\documentclass[12pt]{iopart}
\usepackage{epsfig,graphicx}
\usepackage{amsfonts}
\newtheorem{theorem}{Theorem}
\newtheorem{lemma}[theorem]{Lemma}

\newtheorem{definition}[theorem]{Definition}
\newcommand{\pf}{\noindent \mbox{{\bf Proof}: }}
\newcommand{\blacksquare}{\mbox{\vbox{\hrule\hbox{\vrule\phantom{1}\vrule}\hrule}}}

\begin{document}

 \title[Convergence of the Neumann series for the  Schr\"odinger equation in Banach spaces ]
 {Convergence of the Neumann series for the  Schr\"odinger equation 
and general Volterra equations in Banach spaces
}

 \author{Fernando D. Mera $^1$}

 \address{$^1$ Department of Mathematics, Texas A\&M University,
 College Station, TX, 77843-3368 USA}

\date{September 12, 2011}

 \begin{abstract}

The objective of the article is to treat the 
Schr\"{o}dinger  equation in parallel with a standard treatment of the heat equation. In the mathematics literature, the heat equation initial value problem is converted into a Volterra integral equation of the second kind, and then the Picard algorithm is used to find the exact solution of the integral equation. 
The Poisson Integral theorem shows that the Poisson integral formula with the Schr\"{o}dinger kernel holds in the Abel summable sense. 
Furthermore, the Source Integral theorem provides the solution of the initial value problem for the nonhomogeneous Schr\"{o}dinger equation. Folland's proof of the Generalized Young's inequality is used as a model for the proof of the $L^{p}  $ lemma.  Basically the Generalized Young's theorem is in a more general form where the functions take values in an arbitrary Banach space.  The $L^{1}$, $L^{p}$ and the $L^{\infty}$  lemmas are inductively applied to the proofs of their respective Volterra theorems in order to prove that the Neumman series converge with respect to the topology $L^{p} ( I; \mathcal{B} )$, where $I $ is a finite time interval, $\mathcal{B}$ is an arbitrary Banach space, and $ 1 \leq p \leq \infty $.
The Picard method of
successive approximations is to be used to construct an approximate solution which should approach the exact solution as $ n \to \infty$. To prove convergence, Volterra kernels are introduced in arbitrary Banach spaces. The Volterra theorems are proved and applied in order to show that the Neumann
series for the Hilbert-Schmidt kernel and the unitary kernel converge to the exact Green function. 
  \end{abstract}

\submitto{ISRN Mathematical Physics}
 \maketitle\penalty10000

\section{Introduction} \label{sec:general}
This article is based on a portion of my master thesis \cite{merathesis}. The central problems and theorems of the thesis are summarized in this article . 
The books of the Rubinsteins \cite{Rubinstein}
and Kress \cite{Kress} show how the heat equation is converted to a Volterra integral equation, which is then solved by the Picard algorithm. In this article we shall show that the Schr\"{o}dinger equation has similar properties and results as the heat equation such as the Poisson Integral  Theorem and the Source Integral Theorem. The similarities between
the  Schr\"{o}dinger equation and the heat equation were used to create a theoretical
framework which will give the solution to the Schr\"{o}dinger problem. As much as possible, we use the books \cite{Rubinstein, Kress} as guides to treat the quantum problem like a heat problem. However, 
the parallel between the heat equation and the Schr\"{o}dinger is found to be a limited one, 
and we use the potential theory formalism that Kress laid down in his book in order to study the existence, and uniqueness of the solution of the  Schr\"{o}dinger equation. The
differences between the heat operator and the quantum operator require different
proofs for the uniqueness theorems, the  Source Integral Theorem and the Poisson Integral Theorem.
 For example, the Poisson
integral formula with the Schr\"{o}dinger  kernel is shown to hold in the Abel summable sense.  
In the 
harmonic potential theory, the problem is similarly reduced to an integral equation, but the
integral equation is not of Volterra form and therefore the Neumann series 
does not converge automatically. One can put in a numerical parameter $\beta$ so that the 
series converges when $|\beta| < 1$, but the value one needs for 
the PDE problem 
has $|\beta| = 1$.  In the elliptic PDE problem such as Laplace's equation, Helmholtz equation, or 
Poisson equation, to prove convergence it is not enough to have 
the Banach-space operator having finite norm; the norm must be less than 
unity. The norm must be less than unity, because one needs to bound the norm by the geometric series, not the 
exponential series.  Therefore, the Neumann series is not used to prove 
existence of a solution; instead, the Fredholm theory is used to prove 
existence more abstractly.

The Source Integral Theorem is 
formulated in terms of the source integral term and the initial integral term. The Poisson and the Source Integral theorems for the
Schr\"{o}dinger equation are introduced in section \ref{section1}.
The
initial value problem can be expressed as a Volterra integral equation of the second
kind with respect to time. Our main task is to use the method of successive approximation in
order to prove that there exists a unique solution to the integral equation.
In section \ref{section2}, the article focuses on linear integral operators in arbitrary Banach spaces. 
In section \ref{section3}, the article introduces the Volterra
kernels and applies the Neumann series to give an approximation to the exact
solution.
Folland's proof of the Generalized Young's inequality is used as a  model for the proof of  lemma  \ref{LemmaV3} \cite{Folland}. In section \ref{section3}, we introduce lemma \ref{LemmaV1} and lemma \ref{LemmaV2}  in order to prove the $L^{\infty}$ Volterra theorem and  the $L^{1}$ Volterra theorem, repsectively. Then in each Volterra theorem,  we inductively apply the appropriate $L^{p}$ lemma to prove that the Volterra integral operator can be solved by successive apprximations. In particular, we work with Volterra integral operators $ \hat{Q} $ that go from $ L^{p}(I; \mathcal{B}) $ to itself, where $1 \leq p \leq \infty $. These Volterra integral operators $ \hat{Q} $ are assumed to have uniformly bounded kernels such that $ A : \mathcal{B} \to \mathcal{B}$. Furthermore, we only consider kernels $ A(t, \tau) $ which are Volterra kernels in time. Then the  Volterra theorem proves that Volterra integral equation with a uniform bounded kernel can be solved by successive approximations with respect to the topology $ L^{\infty}(I; \mathcal{B}) $. The $L^{p}$ Volterra theorem  proves the more general case when $ L^{p}(I; \mathcal{B}) $, and where $ 1 < p < \infty$. 
In section \ref{section4}, the article covers two specific kernels,
the Schr\"{o}dinger kernel and the Hilbert-Schmidt kernel. In the Schr\"{o}dinger case, the perturbation expansion series contains a unitary operator and a uniformly bounded potential, and we prove that the Neumann series converges.

\section{The Poisson Integral and Source Integral Theorems}\label{section1}
The wavefunction $\Psi(x,t)$ of a nonrelativistic particle in $\mathbb R^n $ is a solution to the Schr\"{o}dinger equation
\begin{equation}
H\Psi(x,t)=i\hbar \partial_{t} \Psi(x,t)
\end{equation}
where $H$ is the Hamiltonian, given by
\begin{equation}
H = H_{0} + V= \frac{1}{2m} p^{2} + V(x,t) = -\frac{\hbar^{2}}{2m} \Delta_{x} + V(x,t).
\end{equation}
The kinetic operator $ T= \frac{1}{2m} p^{2}$, is also known as the free Hamiltonian $ H_{0}$ in nonrelativistic quantum mechanics. The complex-valued function $\Psi(x,t)$ is the wavefunction, and $|\Psi(x,t)|^{2}$ represents
a particle density function. First, we will consider the case when there is no potential, i.e, $V(x,t)=0.$ Therefore, the free Schr\"{o}dinger equation becomes
\begin{equation} \label{schrodinger}
i   \hbar \partial_{t} \Psi(x,t) = -a^{2} \Delta_{x} \Psi(x,t)
\end{equation}
where 
\begin{equation}
a^{2} = \frac{\hbar^{2} }{2m}.
\end{equation} 
The general abstract Sch\"{o}dinger equation can be realized in several concrete situations such as boundaries, weak potential (where perturbation works), slowly varying potentials (where WKB theory works), and this article is a start on a program dealing with those cases. 

The Schr\"{o}dinger operator and its complex conjugate can be rewritten as 
\begin{equation}
L =a^{2}\Delta_{x} +i\hbar
 \partial_{t}
\end{equation}
and
\begin{equation}
L^{\ast} = a^{2} \Delta_{x}- i \hbar \partial_{t}. 
\end{equation}
Then the adjoint operator of $ L$ is
\begin{equation}
L^{\dagger} = L  = a^{2}\Delta_{x} +i\hbar \partial_{t}
\end{equation}                                                   
with respect to the usual $L^{2}$ inner product and the imposed homogeneous boundary
conditions. In other words the Schr\"{o}dinger operator $L$ is formally self-adjoint.
An important difference to notice is that the heat operator is not formally self-adjoint because the time 
derivative term changes sign.

The nonhomogeneus Sch\"{o}dinger equation with nonhomogeneous initial data is 
\begin{eqnarray}\label{nonSchrodinger}
\eqalign{
L u(x,t) & = a^{2}   \Delta_{x}  u (x, t)  + i \hbar  \partial_{t} u (x,t) = F(x, t) \qquad   \forall (x,t)  \in \mathbb R^{n} \times \mathbb R \cr
u(x, 0 ) & = f(x) \qquad  \forall   (x, t)  \in \mathbb R^{n} \times \{t = 0\}  \cr
}
\end{eqnarray}
where $F(x,t) $ is a source term. In order to get an integral equation for the problem with a potential $V(x,t)$, we will later replace the source term $F(x,t)$  by the potential term $V(x,t) u(x,t) $.

The initial-value problem for the nonhomogeneous Schr\"{o}dinger equation \eref{nonSchrodinger} with nonhomogeneous initial
conditions can be reduced to the analogous problem with homogeneous initial condition by using
the integral representation
\begin{equation}
u(x,t) =  \Phi(x,t) + \Pi(x,t)
\end{equation}                   
where $u(x,t)$ is the solution of the nonhomogeneous problem,  $\Phi(x,t)$ is the source term,  and $\Pi(x,t)$ is the Poisson integral 
term (initial term).  The Poisson integral term and the source term can written as 
\begin{eqnarray}
\Pi(x,t) = e^{- i t H/\hbar}   f(x) =  \int_{\mathbb R^{n} }  K_{f} (x, y, t ) f(y) \, dy 
\end{eqnarray}
and
\begin{eqnarray}
\fl{
\Phi(x, t) = \int_{0}^{t}   e^{- i t H/\hbar} e^{i \tau H/ \hbar}   L u (\cdot, \tau)  \, d\tau =  \int_{0}^{t}  \int_{\mathbb R^{n} }   K_{f} ( x, y, t - \tau)  L u (y, \tau) \, dy d\tau
}
\end{eqnarray}
where $K_{f}(x,y,t)$ is the fundamental solution (free propagator) to the Schr\"{o}dinger equation \eref{schrodinger}  in $\mathbb R^{n} $, and is given by
\begin{equation}
\fl{
K_{f}(x,y,t) \equiv K_{f}(x-y,t) = \biggl( \frac{m}{2\pi \hbar it} \biggr )^{n/2} e^{im |x-y|^{2}/2\hbar  t} \quad  \forall x,y \in \mathbb R^{n}, \, t \neq 0
}
\end{equation}
The free propagator in this case can be expressed in the following manner $K_{f}(x, y, t - \tau)$. The free propagator $K_{f}$ vanishes as a distribution as $t \to 0$ in the region $x \neq y$. In other words,  
the free propagator $K_{f} (x, y, t -\tau)$ satisfies the nonhomogeneous initial condition
\begin{equation}
 K_{f} (x, y, 0) = \lim_{t \to \tau + } K_{f} (x, y, t - \tau) = \delta(x-y) 
\end{equation}
The following theorem introduces the Poisson integral, which is a solution of the
Schr\"{o}dinger equation without a potential term. Our discussion of the Poisson integral is somewhat more 
detailed than that of Evans \cite{Evans}, especially concerning the role 
of Abel summability.
\begin{theorem} \label{theorem1}
Let $f(x)$ be a function on $\mathbb R^{n}$ such that
 $(1+ |y|^{2}) f(y) \in L^{1}(\mathbb R^{n})$. Then the Poisson integral  
\begin{equation} \label{poissonintegral}
u(x,t) = K_{f} \ast f = \int_{\mathbb R^{n}}  K_{f}(x-y,t) f(y) \, dy
\end{equation}
exists in the sense of Abel summability, and is a solution of the equation
\begin{equation}
Lu(x,t) = a^{2} \Delta_{x} u(x,t) + i \hbar  \partial_{t} u(x,t) = 0 \qquad \forall (x,t) \in \mathbb R^{n} \times \mathbb R.
\end{equation}
with given initial data.
The Poisson integral defines a solution of the free Schr\"{o}dinger equation in $\mathbb R^{n}$ ,\,$\forall t \neq 0 $, even $ t<0 $. This solution can be extended into $\mathbb R^{n} \times [0,\infty)$ with the initial condition $u(x,0) = f(x)$ for all points $x$ at which $f$ is continuous. 
\end{theorem}
\goodbreak\pf
If $|y|^{2} f(y) \in L^{1}(\mathbb R^{n})$, then the order of differentiation and integration  in the equation (\ref{poissonintegral}) can be interchanged to show that 
the Poisson integral solves the  Schr\"{o}dinger equation. This hypothesis is obtained from Chapter IV of  \cite{Evans}. 
Let $y=x+\gamma z$, where $\gamma^{2} = \frac{2 \hbar t}{m}$; then we can rewrite the Poisson integral 
as
\begin{equation}
u(x,t)= \biggl (\frac{1}{\pi i }\biggr) ^{n/2}\int_{\mathbb R^{n}} 
e^{i|z|^{2}} f(x+\gamma z)\, dz
\end{equation}
where $|z|=\frac{|x-y|}{\gamma}$.
Let $\epsilon$ be any positive number. Then 
\begin{equation}
(\pi i)^{n/2}  u(x,t)  = \int_{\mathbb R^{n}} 
e^{i |z|^{2}} f(x+\gamma z)\, dz = I_{1} + I_{2} +I_{3},
\end{equation}
where 
\begin{equation} \label{gaussianintegral}
I_{1}  = \int_{\mathbb |z|\leq \epsilon} 
e^{i |z|^{2}} \{ f(x+\gamma z) - f(x) \} \, dz.
\end{equation}
\begin{equation}
I_{2}  =  \int_{\mathbb |z| \geq \epsilon} 
e^{i |z|^{2}} f(x+\gamma z)\, dz
\end{equation}
\begin{equation}
I_{3}  =  \int_{\mathbb |z|\leq \epsilon} 
e^{i |z|^{2}} f(x) \, dz
\end{equation}
Now, we do some calculations for $ I_{1}$ in hyperspherical coordinates $ (\rho, \phi_{1}, \dots, \phi_{n-1}).$ 
The continuity of $f: \mathbb R^{n} \to \mathbb R^{n}$ at some point $y \in \mathbb R^{n}$  implies that $ \forall \, \eta >0 $ $\exists \delta >0$ such that $\forall x \in\mathbb R^{n} $ with $ |x-y| < \delta $ implies that $ |f(x) -f(y) | < \eta $, and $x,y$ are points where $ f  $ is continuous. Given $ \epsilon  $, choose $ \gamma $
such that $ \varepsilon = \gamma \epsilon < \delta $,  and let $ \eta >0 $, then there exists a $ t $ so small such that  $ |f(x + \gamma z) - f(x)|< \eta $ for all $ z$ such that $ |z| \leq \epsilon $.
Therefore, by continuity, we obtain the following bounded estimate:
\begin{equation}
|I_{1} | \leq  \eta \int_{|z| \leq \epsilon}    \, dz
\end{equation}
or
\begin{equation} \label{boundedintegral3}
| I_{1} | \leq   \eta \int_{|z| \leq \epsilon }  \, dz \to 0
\end{equation}
as  $\gamma |z| \to 0$ (i.e., $y \to x$) as $t \to 0$.
Then since, $ f \in L^{1}(\mathbb R^{n})$
\begin{equation}
|I_{2}|  \leq \int_{\mathbb |z|\geq  \epsilon} | f(x+\gamma z)|\, dz \rightarrow 0
\end{equation}
(not necessarily uniformly in $x$) as $\epsilon \to \infty.$ 
In order to handle $I_{3}$ we use the Fresnel integral formula
\begin{equation}
\int_{\mathbb R^{n}} e^{i|z|^{2}}dz= (\pi i)^{n/2}
\end{equation}
A proof of the one-dimensional Fresnel integral formula is outlined on \cite{Carrier}. 
The one-dimensional Fresnel integral implies the product version
\begin{eqnarray} \label{Fresnelintegral} 
\fl{
\int_{\mathbb R^{n}} e^{i|z|^{2}} \, dz = \int_{\mathbb R^{n}} \exp \biggl(i\sum_{k=1}^{n} z_{k}^{2}\biggl) \, dz= \prod_{k=1}^{n} \int_{-\infty}^{\infty} e^{iz_{k}^{2}} \, dz_{k} = \prod_{k=1}^{n} (\pi i)^{1/2} = (\pi i)^{n/2}
}
\end{eqnarray}
Therefore, we have
\begin{equation}
\lim_{\epsilon \to \infty} I_{2} = (\pi i)^{n/2} f(x).
\end{equation}
Then we consider computing the Fresnel integral in terms of polar coordinates instead of Cartesian coordinates. 
Thus, we can rewrite the equation \eref{Fresnelintegral} by
\begin{equation} 
\int_{\mathbb R^{n}} e^{i|z|^{2}} \, dz = \int_{0}^{\infty}   \int_{ \partial B(0,1)}  e^{i \rho^{2}}  \rho^{n-1} \, d\rho d\Omega = \omega_{n} \int_{0}^{\infty} \rho^{n-1} e^{i \rho^{2}} \, d\rho
\end{equation}
Then we make use the substitution $ t = \rho^{2} $, and hence we obtain
\begin{equation} \label{abel}
\int_{\mathbb R^{n}} e^{i|z|^{2}} \, dz  = 
\frac{\omega_{n}}{2}   \int_{0}^{\infty} t^{n-1} e^{i t} \, d t 
\end{equation}
Then we insert the Abel factor $ e^{- \alpha t} $ in the equation \eref{abel} and this gives
\begin{equation}
A(\alpha) = \frac{\omega_{n}}{2}   \int_{0}^{\infty} e^{- \alpha t} t^{n-1} e^{i t} \, d t 
\end{equation} 
and $\omega_{n} $ is the surface area of the unit $n$-sphere. The surface area of the unit $n$-sphere is given by the following formula:
\begin{equation}
\omega_{n} = \frac{2 \pi^{n/2}}{\Gamma (\frac{n}{2}) }
\end{equation}
The complex Gaussian integral (also known as the Gaussian Fresnel integral) can be generalized for any positive integer, i.e., the Gaussian integral is a special case of $ \int_{0}^{\infty} x^{n} e^{-x^{2}} \, dx$,
when $ n=0$. The general Gaussian integral will be shown to be convergent when $n \in \mathbb{N}$: 
Consider the complex integral
\begin{equation}
P_{n} = \int_{0}^{\infty} s^{n-1} e^{i s^{2}} \, ds
\end{equation}
where $ |z|= s= \frac{|x-y|}{\gamma}= \frac{\rho}{\gamma}$. Let $t = s^{2}$,
and substituting this change of variables into equation \eref{complexintegral} we have
\begin{equation}
P_{n} = \frac{1}{2}\int_{0}^{\infty} t^{m} e^{it} \, dt
\end{equation}
where, $m = \frac{n-2}{2}$
and hence,
\begin{equation}  \label{complexintegral}
\int_{0}^{\infty} t^{m}  e^{it} \, dt = \lim_{r \to \infty} \int_{0}^{r} t^{m}    e^{it} \, dt
\end{equation}
Once again, the change of variables $t = iz$ is performed and we have
\begin{equation}  \label{fresnel}
\int_{0}^{r} t^{m}  e^{it} \, dt = i  \int_{0}^{-ir} (iz)^{m}  e^{-z} \, dz
\end{equation}
Then we insert the Abel factor $ e^{-\alpha t}$ into the left-hand side of equation \eref{fresnel} and this gives 
\begin{equation}
\int_{0}^{r} t^{m}  e^{-\alpha t}  e^{it} \, dt = i  \int_{0}^{-ir} (iz)^{m}   e^{- i \alpha z} e^{-z} \, dz
\end{equation}
Then the path of integration is shifted from $0$ to $-i\infty$,
to $0$ to $\infty$. Therefore,
\begin{equation} \label{complexterm}
\lim_{r \to \infty} i \int_{0}^{-ir} (iz)^{m}   e^{-\alpha z} e^{-z} \, dz = \lim_{r \to \infty} i \int_{0}^{r} (iz)^{m}  e^{- \alpha z} e^{-z} \, dz 
\end{equation}
The integral in \eref{complexterm} is convergent for any positive $ \alpha>0$, and the integral over the semicircle at $ \infty $ tends to 0.
In order for the above limit to hold, the path of integration $(iz)^{m}$ must not go over a
branch cut. Thus, if the integrand does not have any poles in the path of integration, then the integral in equation \eref{complexintegral}
is Abel summable and it is related to the gamma function $\Gamma(m+1)$. The integrand in equation \eref{complexterm} is an analytic function, and thus it does not have any poles or branch cuts in the path of integration. In other words,
\begin{equation}
\lim_{r \to \infty} i \int_{0}^{-ir} (iz)^{m}   e^{-  i \alpha z} e^{-z} \, dz =  i^{m+1} \int_{0}^{\infty} z^{m}  e^{-  i \alpha z} e^{-z} \, dz 
\end{equation}
Then, we take the limit of $\alpha \to 0$, and the above equation becomes
\begin{equation}
\lim_{\alpha \to 0} i^{m+1} \int_{0}^{\infty} z^{m}  e^{- i \alpha z} e^{-z} \, dz  =   i^{m+1}  \int_{0}^{\infty} z^{m} e^{-z} \, dz 
\end{equation}
and this limit holds in the Abel sense, and we take $i^{m} = e^{i(\pi/2)m}$. Thus,
\begin{equation}
P_{n} = \frac{1}{2}i^{m+1} \Gamma(m+1) = \frac{i^{n/2}}{2} \Gamma \biggl(\frac{n}{2} \biggr)
\end{equation}
and hence,
\begin{equation}
A(0) = \int_{\mathbb R^{n}} e^{i |z|^{2}} \, dz =   \frac{i^{n/2} \omega_{n} }{2} \Gamma \biggl(\frac{n}{2} \biggr)
\end{equation}
This confirms equation \eref{Fresnelintegral} in an alternative way. 
 
This implies the continuity of $u(x,t)$ at $t = 0$. Therefore, the Poisson integral has the initial values $u(\cdot, 0) = f(x)$ for all points $x$  at which $f$ is continuous. \qquad $\square$\\
The proof of theorem \ref{theorem2} is partly based on \cite{Hagedorn}. In \cite{Hagedorn}, the authors construct an abstract lemma for a function $\psi(t)  $ that belongs to the domain of the operator $H$, and approximately solves the Schr\"{o}dinger equation in the following sense
\begin{eqnarray}
i \hbar \frac{\partial  \psi(t) }{ \partial t } = H \psi(t) + F(t) 
\end{eqnarray}
where $F(t)$ is a source term. In our case, we want to prove that there exists a function $u(t)$ that is the exact solution to the same Schr\"{o}dinger problem with the difference that $H =H_{0} = - a^{2} \Delta $. The other difference is that we are trying to prove that the solution can be expressed a Volterra integral equation of the second kind.  

All throughout this article the letter $I $ will denote the time interval $(0,T)$, where $T$ is a positive constant. In theorem \ref{theorem2}, we will define the $ L^{\infty} (I; \mathcal{B})$ norm of a function $ u $ to be
\begin{eqnarray}
\fl{
\eqalign{
\| u \|_{L^{\infty} (I ; \mathcal B)  }  =  \inf \{ M \geq 0 : \| u(t) \|_{\mathcal{B}}  \leq M, \mbox{ holds for almost all $t \in [0,T]$ } \}
}
}
\end{eqnarray}
where $\mathcal B$ is a Banach space. For a more formal and general definition of the $L^{\infty}$ norm look at definitions \eref{DefinitionBanach}-\eref{DefinitionBanach2}. 
\begin{theorem}\label{theorem2}
Let $f(x)$ be a function on $\mathbb R^{n}$ with the following property: $(1+ |y|^{2}) f(y) \in L^{1}(\mathbb R^{n})$. Furthermore, suppose the source term $F(x,t) $ satisfies the condition
\begin{eqnarray}
\| F(\cdot ,t) \| \leq \xi(t) 
\end{eqnarray}
where $\| \cdot \|$ is a Banach space norm
and the function $\xi(t)$ satifies the following condition
\begin{equation}
\| \xi \|_{L^{\infty} (I )  } \leq M 
\end{equation}
where  $M$ and $ T$ are positive constants.   The solution of the initial-value problem for the nonhomogeneous Schr\"{o}dinger equation \eref{nonSchrodinger} can be represented as the following 
integral formula:
\begin{equation}
u(x,t) =  \Phi(x,t) + \Pi(x,t)
\end{equation} 
The initial term,  and the source term  are given by the following integral representation formulas:
\begin{equation}
\Pi(x,t) = \int_{\mathbb R^{n} } 
K_{f}(x, y , t)  f(y)  \, dy
\end{equation}
and
\begin{equation} \label{tildeK}
\Phi(x,t) =  - \frac{ i  }{\hbar } \int_{0}^{t} \int_{\mathbb R^{n} } K_{f}(x,t;y,\tau) F(y,\tau) \,dy d\tau 
\end{equation}                             
where,  $K_{f}(x,t;y,\tau)$ is the fundamental solution and $u(x, 0)= f(x)$. The solution $u$ will belong to the Banach space $L^{\infty}(I; \mathcal{B})$.
\end{theorem}
\goodbreak\pf
Suppose that $f(x)$ be a function on $\mathbb R^{n}$ with the following property: $(1+ |y|^{2}) f(y) \in L^{1}(\mathbb R^{n})$. Let us assume $t >0$. The Poisson integral term was shown to solve the initial-value problem for the homogeneous Schr\"{o}dinger equation. Let $u(x,t)$ be the solution to the nonhomogeneous Schr\"{o}dinger equation \eref{nonSchrodinger}. Then we claim that solution $u(x,t)$ can be expressed as a Volterra integral representation
\begin{eqnarray}\label{Unitaryop}
\fl{
 u(x,t)      =  \int_{\mathbb R^{n} } 
K_{f}(x, y , t)  f(y)  \, dy  - \frac{ i  }{\hbar } \int_{0}^{t} \int_{\mathbb R^{n} } K_{f}(x,t;y,\tau) F(y,\tau) \,dy d\tau  
}
\end{eqnarray}
In other words, we claim that the solution for the nonhomogeneous Schr\"{o}dinger equation is given by
\begin{eqnarray}
u(x, t )=  u_{h} (x, t) + u_{p}(x, t)
\end{eqnarray}
where $u_{h}$ is  the solution of the homogeneous Schr\"{o}dinger equation, and $u_{p}$ is the particular solution of the nonhomogeneous Schr\"{o}dinger equation. Then  by the Poisson integral theorem, the function $u_{h}$ is the solution of the homegenous Schr\"{o}dinger equation and hence $L u_{h} = 0$, $\forall (x, t) \in \mathbb R^{n} \times \mathbb R$. Then by applying the Schr\"{o}dinger operator to $u(t) $, we have 
\begin{eqnarray}
\fl{
\eqalign{
L u & = L u_{p} = - a^{2}  \Delta_{x}  u_{p} + i \hbar \frac{ \partial  u_{p}  }{\partial t }  = \\
& = - a^{2}  \int_{\mathbb R^{n} } 
 \Delta_{x}   K_{f}(x, y , t)  f(y)  \, dy  +   i \hbar \frac{ \partial    }{\partial t }   \biggl ( - \frac{i  }{ \hbar}  \int_{0}^{t} \int_{\mathbb R^{n} } K_{f}(x, y,t - \tau) F(y,\tau) \,dy d\tau  \biggr ) \\
}
}
\end{eqnarray}
but $\Delta_{x}   K_{f} (x, y, t - \tau) = 0 $, $\forall t > \tau$, and hence
\begin{eqnarray}
\eqalign{
L u & =  L u_{p}= \frac{ \partial    }{\partial t }   \biggl ( \int_{0}^{t} \int_{\mathbb R^{n} } K_{f}(x, y,t - \tau) F(y,\tau) \,dy d\tau  \biggr) \\
&
= \int_{\mathbb R^{n} } K_{f}(x, y,t - t)  F(y,t ) \,dy =  \int_{\mathbb R^{n} } K_{f}(x, y,0)  F(y,t ) \,dy \\
& =  \int_{\mathbb R^{n} } \delta(x - y ) F(y,t ) \,dy = F(x,t)
}
\end{eqnarray}  
and hence the solution $u(t)$ is given by the Volterra integral \eref{Unitaryop}.

Now, this equation should satisfy the initial condition $u(x, 0) = f(x)$. This implies that the second term should vanish as $t \to 0$. 

Another way to express equation  \eref{Unitaryop} is via unitary operators. Hence,
\begin{eqnarray}
 e^{- i t H/\hbar}   f(x) - \frac{ i }{\hbar}  \int_{0}^{t}   e^{- i t H/\hbar} e^{i \tau H/ \hbar}  F (\tau)  \, d\tau =  u(t)    
\end{eqnarray} 
Let us consider the following integral,
\begin{eqnarray}
- i \hbar^{-1} \int_{0}^{t}  e^{i \tau H / \hbar}  L u (\tau) \, d\tau   
\end{eqnarray}
and since $L u  = - a \Delta u +  i \hbar \frac{\partial  u }{\partial \tau} $, we can rewrite the above equation in the following way
\begin{eqnarray}
\fl{
\eqalign{
- i \hbar^{-1} \int_{0}^{t}  e^{i \tau H / \hbar}  L u (\tau) \, d\tau    & = - i \hbar^{-1} \int_{0}^{t}   e^{i \tau H / \hbar}  \biggl ( 
- a \Delta u(\tau)  +  i \hbar \frac{\partial  u(\tau) }{\partial \tau}  \biggr ) \, d\tau \\
&
 = - i \hbar^{-1} \int_{0}^{t}   e^{i \tau H / \hbar}  \biggl ( - H u(\tau) + i \hbar \frac{\partial  u }{\partial \tau}  \biggr ) \, d\tau \\
 & 
 = \int_{0}^{t}  \frac{\partial  }{\partial \tau}   \biggl (  -  e^{i \tau H / \hbar}     u(\tau)   \biggr ) \, d\tau 
= 
 -   e^{i t H / \hbar}  u(t) + u(0) \\
 & = u(0)  -   e^{i t H / \hbar}  u(t) 
}
}
\end{eqnarray}
Then, we can rewrite the above equation into the following form:
\begin{equation}
u(t) =   e^{  - i t H / \hbar}    u(0) - i \hbar^{-1}    e^{  - i t H / \hbar}   \int_{0}^{t}  e^{i \tau H / \hbar}  L u (\tau) \, d\tau
\end{equation}
and this is the Volterra integral equation \eref{Unitaryop} written in unitary operator notation.
Let us take expression $ e^{- i t H/\hbar}   u(0)  - u( t)$ into consideration. This expression is simply the  source term $\Phi(x,t)$. Then we take a Banach space norm to the quantity $ e^{- i t H/\hbar} u(0)  - u(t)$, and  by the unitarity of the propagator and the fundamental theorem of calculus, we have
\begin{eqnarray}
\fl{
\eqalign{
\| \Phi (t) \| & = \|  e^{- i t H/\hbar}   u(0)  - u(t) \|  
 =
 \biggl  \|    i \hbar^{-1}   e^{i t  H/ \hbar}  \int_{0}^{t}    e^{i \tau  H/ \hbar}  L u(\tau) \, d\tau \biggr \| \\
 & \leq h^{-1} \int_{0}^{t}  \| Lu \| \, d\tau   \leq   \hbar^{-1 } \int_{0}^{t}  \xi(\tau)  \, d\tau 
}
}
\end{eqnarray}
since by hypothesis we have that $\| Lu \| \leq \xi(\tau) $.
Hence,
\begin{eqnarray}
\fl{
\eqalign{
\| \Phi(t) \| = \|  e^{- i t H/\hbar}   u(0)  - u(t) \|  &\leq \hbar^{-1 } \int_{0}^{t}  \| \xi  \|_{L^{\infty}( I  ) }  \, d\tau\\
& \leq 
 \hbar^{-1 }  \| \xi  \|_{L^{\infty}(  I ) }   t
}
}
\end{eqnarray}
Then by taking the limit of $t \to 0$, the above inequality also goes to $0$.  The Poisson integral term $\Pi(x,t ) $ satisfies the initial condition by theorem \ref{theorem1}. Therefore, the Volterra integral equation \eref{Unitaryop} satisfies the initial condition and is a solution to the nonhomogeneous Schr\"{o}dinger equation. \qquad  $\square$
\section{Integral Equations and Neumann Series}\label{section2}
In this section, we introduce the integral operators in arbitrary Banach spaces
in order to find a solution to the Schr\"{o}dinger equation in $\mathbb R^{n+1}$. This section is an
informal preview of the Volterra and General Volterra Theorems which will be proved in section \ref{section3}.
In the following analysis of integral operators, this article will use as a foundation
 Kress's treatment of linear integral equations \cite{Kress}.
In operator notation, the Volterra integral
equation of the second kind is written in the following manner:
\begin{equation}\label{Volterraintegral}
\phi-\hat{Q}\phi=f
\end{equation}
where it is assumed that $ \hat{Q}$ is a bounded linear operator from a Banach space $ \mathcal{B}$ to itself and  $\phi , f \in \mathcal{B}$. The existence and uniqueness
of a solution to an integral operator equation can be found via the inverse
operator $(I-\hat{Q})^{-1}$, where $I$ is the identity operator. The existence of the inverse operator will become clear below.
\begin{definition} 
Let $BL(\mathcal{B};\mathcal{B})$ be the collection of bounded linear transformations from
$\mathcal{B}$ into $\mathcal{B}$. Also, we denote the space $BL(\mathcal{B}, \mathbb{F})$ as the set of bounded linear functionals on $\mathcal{B}$, where $ \mathbb F = \{\mathbb R, \mathbb C \} $.
\end{definition}
Important Banach spaces which we will be dealing with are the Lebesgue spaces $
L^{p}(\mu)$. This article will cover the case when $p = \infty$, in order to create bounded
estimates of the Volterra operator $\hat{Q}$ with respect to the norm $\| \cdot \|$. In the next
section, the Volterra Theorem will prove that the spectral radius of the
Volterra operator is zero using the $L^{\infty}$-estimates.
\begin{definition}\label{DefinitionBanach}
Let $(\Omega,\Sigma,\mu)$ be a measure space 
and $\mathcal{B}$ be a Banach space. The collection of all essentially bounded measureable functions on 
$\Omega$ taking values in ${\mathcal B}$ is denoted 
$L^\infty(\Omega,\mu;{\cal B})$, and the reference to $\mu$ is 
omitted when there is no danger of confusion. The essential supremum of a function $\varphi \colon \Omega \to 
\mathcal{B}$ is given by
\begin{eqnarray}
 \|\varphi\|_{L^\infty (\Omega;{\cal B})} =
 \inf\{M\ge 0: \|\varphi(x)\|_{\mathcal{B} \to \mathcal{B} } \leq M \, \mbox{holds for almost all $x$}\}. 
\end{eqnarray}
If $ \varphi $ does have an essential bound, then it is said to belong to  $L^\infty(\Omega;{\cal B})$.
\end{definition}
\begin{definition}\label{DefinitionBanach2}
 Let $\mathcal{B}_1$ and $\mathcal{B}_2$ be Banach spaces and $\Omega$ be some measurable space.
  Suppose that  the function $A(\cdot,\cdot)$ is  measurable, and define 
 its norm
\begin{equation}
\| A(x,y) \|_{\mathcal{B}_{1}  \to \mathcal{B}_{2} } =  \inf\{ M \geq 0 : \|A(x,y) \phi \| \leq M \| \phi \|, \forall \phi  \in \mathcal{B}_{1} \}
\end{equation}
 If $\mathcal{ B  }_{1}     = \mathcal{B}_{2} = \mathcal{B}$, then we use the following notation for the Banach space norm of $A(x, y)$:
\begin{eqnarray}
\| A(x,y) \| = \| A(x,y) \|_{\mathcal{B}  }  =  \inf\{ M \geq 0 : \|A(x,y) \phi \| \leq M \| \phi \|, \forall \phi  \in \mathcal{B} \}
\end{eqnarray}
For each $(x,y)\in \Omega \times \Omega$ let $A(x,y) : \mathcal{B}_1 \to \mathcal{B}_2$ be a bounded linear 
operator with norm
\begin{eqnarray}
\| A \|_{L^{\infty}(\Omega^{2}  ; \mathcal{B}_{1}  \to  \mathcal{B}_{2}   ) } & 
 \equiv \inf \{ M \geq 0 :\,  \| A(x,y) \| \leq M, 
\mbox{for almost all }(x ,y)\in \Omega^{2} \}  \nonumber \\
& =\mathrm{ess\,sup}_{(x,y) \in \Omega^{2}} \| A (x,y)\|_{\mathcal{B}_{1}  \to  \mathcal{B}_{1} } 
\end{eqnarray}
and  $A(\cdot,\cdot)$ is called a bounded operator kernel if 
  $\|A\|_{L^\infty(\Omega^2; \mathcal{B}_{1}  \to \mathcal{B}_{2}  ) }$ is finite. 
\end{definition}
In Section \ref{section1} we have reformulated the Schr\"odinger equation as an integral equation of the second 
kind.  The existence and uniqueness of its solution can be found by analysis of the Neumann series.
The successive approximations
\begin{equation}\label{approx}
\phi_{n+1} = \hat{Q} \phi_{n} +f
\end{equation}
converge to the exact solution of the integral equation \eref{Volterraintegral}, if some technical conditions are satisfied. Otherwise, 
the successive approximations may not converge even for the Volterra operator, e.g. if its kernel is not good enough or the interval is infinite.

The above integral equations are given for an arbitrary
Banach space $\mathcal{B}$ that will be used in Picard's algorithm of successive approximation.
Then, equation \eref{approx} is converging to the solution $\phi$ if the following conditions are
satisfied: \\
1)the integral operator $\hat{Q}$ is a bounded linear operator in the Banach
space $\mathcal{B}$. \\
2)the function $f$ belongs to a Banach space $\mathcal{B}$, \\
3)and finally, the infinite series $ \varphi =  \sum_{j=0}^{\infty} \hat{Q}^{j} f$ is a convergent series  with respect to the topology of $ L^{\infty} $ in time and  of $ \mathcal{B} $ in space. 

If these three conditions are satisfied, then it is possible to use the Neumann series
to obtain the exact solution to the original problem, which is the initial 
value problem of the Schr\"{o}dinger equation with a potential term $V(x,t)$. The three conditions turn out be the necessary hypotheses to prove the Volterra and $L^{p}$ Volterra theorems.
\section{Volterra Kernels and Successive Approximations}    \label{section3}               
In this section we will revisit the method of successive approximations. 
Our main theorem will prove the well-known fact that the Volterra integral operator of the second
kind has a spectral radius of zero. This direct proof is not found in \cite{Kress},  where the logic runs in the other direction. 
That is, our proof of the Volterra and General Volterra theorems will not use the spectral radius to prove that the Neumann series converges. The Volterra operator is known to have a nice property, known as the simplex structure. It is the simplex structure which make the infinite Neumann series converge.  Then it follows from the convergence of the Neumann series that the spectral radius is zero.

Now, we focus our attention to four hypotheses which will be used in lemma \ref{LemmaV1} and lemma \ref{LemmaV2} . Lemma \ref{LemmaV1} will be inductively applied  in proving the $L^{\infty} $ Volterra theorem. Similarly, lemma \ref{LemmaV3} will also be inductively applied to the $L^{p}$ Volterra theorem. The following are the four hypothesis that are needed for lemmas \ref{LemmaV1}, \ref{LemmaV2}, and \ref{LemmaV3} :\\
1.) Let $\mathcal{B}$ be a Banach space, and let $I$ be an interval. \\
2.) Assume that $ \forall (t, \tau) \in \bar{I}^{2}$, $A(t, \tau): \mathcal{B} \to \mathcal{B}$.\\
3.)  Suppose that  the kernel $A(t,\tau)$ is a measurable and uniformly bounded linear operator (in the sense of definition 6) .\\
4.) Finally, let $A(t, \tau) $ satisfy the Volterra condition, $A(t, \tau) = 0$ if $\tau >t $. 

In theorem \ref{volterratheorem}, we have that $\Omega=I= (0, T)$, and the Lebesgue space is $L^{\infty} (I; \mathcal{B} )  $, i.e. $p = \infty$. And for theorem \ref{L1Volterra}, $\Omega = I$, and the Lebesgue space we consider is $L^{1}(I ; \mathcal{B} )$, where $ p  = 1$. Lemma \ref{LemmaV1} and lemma \ref{LemmaV2} are needed to conclude that $\|  \hat{Q} \phi \|_{L^{p } (I; \mathcal{B} ) }  \leq D t \| \phi \|_{L^{p}( I; \mathcal{B}) }$ for  $D >0$ and $t \in [0,T]$,  where $p = 1$ and  $p = \infty$. Finally, lemma \ref{LemmaV3} is needed in theorem \ref{LpVolterra} to conclude that $\|  \hat{Q} \phi \|_{L^{p } (I; \mathcal{B} ) }  \leq  M(p )  t\| \phi \|_{L^{p}( I; \mathcal{B}) }$ where  $M(p)  = \frac{D}{ p^{1/p}  }$ , $t \in [0,T]$ and $1< p < \infty$. In the lemmas \ref{LemmaV1}-\ref{LemmaV3}, $n$ is any
real positive number.
\begin{lemma}\label{LemmaV1}
 Let the Volterra integral operator, $\hat{Q}: L^{\infty}(I;\mathcal{B} ) \to L^{\infty}( I ; \mathcal{B})$, be defined by
\begin{equation}
\hat{Q}\phi(t) = \int_{0}^{T} A(t,\tau) \phi(\tau ) \, d\tau = \int_{0}^{t} A(t,\tau)\phi(\tau) \, d\tau ,
\end{equation}
where $\phi \in \mathcal{B}$. Let $\phi \in L^{\infty}(I; \mathcal{B}  )$, and assume that $\exists C>0$ , for each subinterval $J$ of the form $(0,t)$, such that $\| \phi \|_{L^{\infty}( J ; \mathcal{B}  ) }  =\sup_{    0 < \tau < t}   \| \phi(\tau )  \|    \leq  C  t^{n}$. Assume that hypotheses (1)-(4) are satisfied. Then it follows that $\| \hat{Q}  \phi\|_{L^{\infty }  ( J  , \mathcal{B}  ) }\leq \frac{DC t^{n+1}  }{n+1}$.
\end{lemma}
\goodbreak\pf 
Since $A(t, \tau) $ satisfies definition 6, then  $\exists D>0$ such that  $\forall (t, \tau) \in \bar{I}^{2}$,  $\| A(t, \tau) \| \leq D < \infty $. The $L^{\infty}(J  ;\mathcal{B}  )$ norm of the function $\hat{Q} \phi(t)$ is
\begin{eqnarray}
\eqalign{
\| \hat{Q}  \phi  \|_{L^{\infty} (I_{t} ; \mathcal{B}  ) } & = \sup_{t_{1}  \leq t } \biggl \|   \int_{0}^{t_{1}  } A(t, \tau) \phi(\tau)  \, d\tau  \biggr \|   \leq\sup_{t_{1}  \leq t }  \int_{0}^{t_{1}  }   \|   A(t, \tau) \phi(\tau)  \| \, d\tau \\
&  \leq \sup_{  t_{1}  \leq t }  \int_{0}^{ t_{1} } \| A( t, \tau) \| \, \| \phi(\tau ) \| \, d\tau \leq \sup_{  t_{1}  \leq t }  \int_{0}^{t_{1}  }  D C \tau^{n} \, d\tau  \\
& = \sup_{  t_{1}  \leq t }  D C \frac{  t_{1} ^{n + 1}   }{ n+ 1}  = \frac{D C t^{n+ 1}   }{ n+ 1 } 
}
\end{eqnarray}
\qquad $\square$
\begin{lemma}\label{LemmaV2}
Let the Volterra integral operator, $\hat{Q}: L^{1}(I;\mathcal{B} ) \to L^{1}( I ; \mathcal{B})$, be defined by  
\begin{equation}
\hat{Q}\phi(t) = \int_{0}^{T} A(t,\tau) \phi(\tau ) \, d\tau = \int_{0}^{t} A(t,\tau)\phi(\tau) \, d\tau ,
\end{equation}
where $\phi \in \mathcal{B}$. Let $\phi \in L^{ 1 }(I; \mathcal{B}  )$,  and assume that $\exists C>0$, for each subinterval $J$ of the form $(0,t)$, such that 
\begin{eqnarray}
\| \phi \|_{L^{ 1 }( J ; \mathcal{B}  ) }   = \int_{0}^{t}   \| \phi(\tau )  \|  \, d\tau    \leq  C  t^{n}
\end{eqnarray}
Assume that hypotheses (1)-(4) are satisfied.  
 Then it follows that $\| \hat{Q}  \phi\|_{L^{ 1 }  ( J  , \mathcal{B}  ) }\leq \frac{DC t^{n+1}  }{n+1}$.
\end{lemma}
\goodbreak\pf 
Since $A(t, \tau) $ satisfies definition 6, then  $\exists D>0$, such that  $\forall (t, \tau) \in \bar{I}^{2}$,  $\| A(t, \tau) \| \leq D < \infty $. The $L^{1}(J  ;\mathcal{B}  )$ norm of the function $\hat{Q} \phi(t)$ is
\begin{eqnarray}
\eqalign{
\| \hat{Q}  \phi  \|_{L^{1} ( J ; \mathcal{B}  ) } & = \int_{0}^{t}  \biggl \|   \int_{0}^{t_{1}  } A(t, \tau) \phi(\tau)  \, d\tau  \biggr \|  \, d t_{1} 
\leq  \int_{0}^{t}  \int_{0}^{t_{1}  }   \|   A(t, \tau)  \| \, \| \phi(\tau)  \| \, d\tau \\
&  \leq  \int_{0}^{t}  D    \int_{0}^{ t_{1} }  \| \phi(\tau ) \| \, d\tau d t_{1} 
\leq  \int_{0}^{t}  D \| \phi \|_{L^{1}( J ; \mathcal{B}  ) }  \,   d t_{1}   \\
& \leq \int_{0}^{t}   D C t_{1} ^{n} \,   d t_{1}   = \frac{D C t^{n+ 1}   }{ n+ 1 } 
}
\end{eqnarray}
\qquad $\square$\\
The  proof of  lemma \ref{LemmaV3} uses Folland's proof of Young's inequality as a model\cite{Folland}. 
\begin{lemma}\label{LemmaV3}
Let the Volterra integral operator, $\hat{Q}: L^{ p }(I;\mathcal{B} ) \to L^{ p }( I ; \mathcal{B})$, be defined by
\begin{equation}
\hat{Q}\phi(t) = \int_{0}^{T} A(t,\tau) \phi(\tau ) \, d\tau = \int_{0}^{t} A(t,\tau)\phi(\tau) \, d\tau ,
\end{equation}
where $\phi \in \mathcal{B}$.  Let $\phi \in L^{p}(I; \mathcal{B}  )$, where $ 1< p < \infty$, and assume that for each subinterval $J$, $\exists C>0$ such that
\begin{eqnarray}
\| \phi \|_{L^{ p }( J ; \mathcal{B}  ) }   = \biggl ( \int_{0}^{t }   \| \phi(\tau )  \|^{p}  \, d\tau \biggr )^{1/p}    \leq  C t^{n} .
\end{eqnarray}
Assume that hypotheses (1)-(4) are satisfied.
Then it follows that $\| \hat{Q}  \phi\|_{L^{ p  }  ( J  , \mathcal{B}  ) }\leq C D \frac{ t^{ n +1}  }{ [  p( n  +1 )]^{1/p} }$.
\end{lemma}
\goodbreak\pf 
Suppose  $1< p < \infty$ , and let $q $ be the conjugate exponenet ($ p^{-1} + q^{-1}  = 1  $).  Since $A(t, \tau) $ satisfies definition 6, then  $\exists D>0$, such that  $\forall (t, \tau) \in \bar{I}^{2}$,  $\| A(t, \tau) \| \leq D < \infty $. The Banach space norm of the function $\hat{Q} \phi(t)$ is
\begin{eqnarray}
\fl{
\eqalign{
\|  \hat{Q}  \phi( t_{1} ) \|  & \leq  \biggl  (  \int_{0}^{t_{1} }  \| A(  t_{1}, \tau)  \| ) \,   d\tau   \biggr )^{1/q}  \biggl ( \int_{0}^{t_{1}  }  \| A(t_{1}  , \tau) \| \, \| \phi( \tau ) \|^{p}  \, d\tau \biggl )^{1/p}  \\
& \leq D^{1/q} \biggl  (  \int_{0}^{t_{1} }      d\tau   \biggr )^{1/q}  \biggl ( \int_{0}^{  t_{1}  }  D \|  \phi( \tau ) \|^{p}  \, d\tau \biggl )^{1/p} 
\leq D^{1/q}   D^{1/p}   t_{1}^{1/q}  \biggl ( \int_{0}^{  t_{1}  }  \| \phi ( \tau ) \|^{p}   d\tau \biggl )^{1/p}  \\
& 
\leq D t_{1 }^{1/q} \biggl ( \int_{0}^{  t_{1}  }  \| \phi ( \tau ) \|^{p}  \, d\tau \biggl )^{1/p}  
}
}
\end{eqnarray}
and then we raise both sides to the $p$th power and integrating, we see by Fubini's theorem that
\begin{eqnarray}
\fl{
\eqalign{
 \int_{0}^{  t_{1}  }   \| \hat{Q}  \phi( t_{1})  \|^{p}  \, d t_{1}  &  \leq  \int_{0}^{   t }   D^{p}  t_{1}^{p/q}  \int_{0}^{  t_{1}  }  \| \phi ( \tau ) \|^{p}  \, d\tau d t_{1}  
\leq  \int_{0}^{   t }    D^{p}   \int_{0 }^{ t  }      t_{1}^{p/q}  \| \phi  \|_{  L^{p }  ( J ; \mathcal{ B}  ) }^{p}  \,  d t_{1}   \\
&
 \leq D^{p}    \int_{0}^{t}    C   t_{1}^{n p  + p/q } \, d t_{1}  
  \leq D^{p} C^{p}   t^{p/q}  \frac{ t^{n p + p/q + 1  }  }{  n p + \frac{p }{q } + 1 }   \\
 &  \leq D^{p}  C^{p}     \frac{ t^{n p + p/q  + 1  }  }{  n p + p  }  
}
}
\end{eqnarray}
since $\frac{  1 }{ q }  = 1 - \frac{  1 }{ p }$, and hence $\frac{ p }{ q }  = p - 1$. Now we take the $p$-th root, and this yields
\begin{eqnarray}
\fl{
\| \hat{Q}  \phi  \|_{  L^{p }  ( J ; \mathcal{ B}  ) }  \leq   DC \frac{ t^{n  + 1/p + 1/q}   }{ [   p( n  + 1)  ]^{1/p} }    = D C    
\frac{ t^{n  + 1 }  }{ [  p  (n  + 1)  ]^{1/p} }  =  DC \frac{ t^{n  + 1 }  }{ [   p ( n+ 1)  ]^{1/p} }  
}
\end{eqnarray}
\qquad $\square$

The following theorem is the main theorem of \cite{merathesis}. It is also used in the examples of section \ref{section3}. The General Volterra Theorem is s just a variant of the Volterra Theorem, i.e., it is the $L^{p}$-analogue.
\begin{theorem}\label{volterratheorem} ($L^{\infty}$ Volterra Theorem) Let $I = (0, T) $ such that $0 < T < \infty$. Assume that
the kernel $A(t,\tau)$ is a measurable and uniformly bounded linear operator. Furthermore, assume that
$\forall (t, \tau) \in \bar{I}^{2} $,  $A(t, \tau): \mathcal{B} \to \mathcal{B}$ where $\mathcal{B}$ is a Banach space. Suppose that the kernel satisfies the following condition, $A(t,\tau ) = 0$, when $ \tau > t.$  The Volterra integral operator, $\hat{Q}: L^{\infty}(I;\mathcal{B} ) \to L^{\infty}(I ; \mathcal{B})$, is defined by         
\begin{equation}
\hat{Q}\varphi(t) = \int_{0}^{T} A(t,\tau) \varphi(\tau ) \, d\tau = \int_{0}^{t} A(t,\tau)\varphi(\tau) \, d\tau ,
\end{equation}
where $\varphi \in \mathcal{B}$.
Then, the Volterra integral equation with the kernel $A(t,\tau)$ can be solved by successive approximations. That is, the
Neumann series converges in the topology of $L^{\infty}(I;\mathcal{B} )$.
\end{theorem}
\goodbreak\pf
Let $  L^{\infty}( J  ;\mathcal{B}  )$ be the Banach space with norm $ \| \cdot \|_{ L^{ \infty }( J ; \mathcal{ B }  )  }$, where $J = (0, t )$ and  where $0 < t \leq T$.  Assume that hypotheses (1)-(4) are satisfied. Then by definition 6, $\exists D>0$ such that $\forall (t, \tau) \in  \bar{I}^{2}$, $\| A(t, \tau) \| = D < \infty$.

Suppose the function $ \psi_{0} : J \to L^{\infty} (J  ;  \mathcal{B}  )$ is a bounded function with norm $\| \psi_{0} \|_{L^{ \infty }( J ;\mathcal{ B }   ) } = \sup_{ 0 < \tau < t}    \|\psi_{0}(\tau) \|   $. Then by lemma \ref{LemmaV1}, it follows that the norm of $\psi_{0} (t) $ satifies the following condition, $\| \psi_{0} \|_{L^{\infty}( J ;\mathcal{B} )}  \leq N t^{0} = N$, since $n = 0$. Then, we compute $L^{\infty}$ norm estimates for the following equation:
\begin{eqnarray}
\eqalign{
\| \psi_{1}  \|_{L^{\infty}( J ;  \mathcal{B} ) } &= \|\hat{Q} \psi_{0} \| _{L^{\infty}( J;  \mathcal{B} ) }  = 
\sup_{t_{1} \leq t}  \biggl \| \int_{0}^{t_{1}   }  A( t_{1}, \tau ) \psi_{0}  (\tau) \, d\tau   \biggr \| \\
& \leq 
\sup_{t_{1} \leq t}   \int_{0}^{t_{1} } \|A(t,\tau)\| \, \|\psi_{0}(\tau) \| \, d\tau \nonumber  \leq \sup_{t_{1} \leq t}   \int_{0}^{t_{1} }   D N  \,d\tau    = D N t 
}
\end{eqnarray}
and if $ N = \| \psi_{0}  \|_{L^{\infty}(  J ;  \mathcal{B} ) }$, then if follows that 
\begin{equation} \label{bdestimate}
\|\hat{Q} \psi_{0} \| _{L^{\infty}( J ;  \mathcal{B} ) }    \leq   D  \| \psi_{0} \| _{L^{\infty}(  J ;  \mathcal{B} ) }  t.
\end{equation}
Then by inductively applying lemma \ref{LemmaV1}, we see that the nth term of the Neumann series $\psi_{n}$ gives the simplex structure:
\begin{equation}
 \|\psi_{n}  \|_{L^{\infty }(  J ; \mathcal{B}  ) }  \leq    D^{n}\| \psi_{0} \|_{L^{\infty}(J ; \mathcal{B})  } \frac{t^{n}}{n!}
\end{equation}
Therefore the series $ \sum_{n=0}^{\infty}  \psi_{n}$ is majorized by
\begin{eqnarray}
\fl \| \psi_{0} \|_{L^{\infty}(J ; \mathcal{B})} + 
\| \psi_{0} \|_{L^{\infty}( J; \mathcal{B})} \sum_{n=1}^{\infty}
 \frac{   D^{n}    t^{n}}{n!}
 & = \| \psi_{0} \|_{L^{\infty}(J; \mathcal{B})} \sum_{n=0}^{\infty}
 \frac{  D^{n}  t^{n}}{n!} 
\nonumber \\
& = \| \psi_{0} \|_{L^{\infty}(J ; \mathcal{B})} e^{  D  t}.
\end{eqnarray}
and the above estimate holds for all $t \in (0, T]$. 
Therefore, the Neumann series converges in the topology of $L^{\infty}(I;\mathcal{B} )$. \quad $\square$

\begin{theorem} \label{L1Volterra}($L^{1}$ Volterra Theorem)
 Let $I = (0, T) $ such that $0 < T < \infty $. Assume  that the kernel $A(t,\tau)$ is a measurable and uniformly bounded linear operator(in the sense of definition 6).  Furthermore, assume that
$\forall (t, \tau) \in \bar{I}^{2} $,  $A(t, \tau): \mathcal{B} \to \mathcal{B}$ where $\mathcal{B}$ is a Banach space. Suppose that the kernel satisfies the following condition, $A(t,\tau ) = 0$, when $ \tau > t$.  The Volterra integral operator, $\hat{Q}: L^{ 1 }( I;\mathcal{B} ) \to L^{1}(I ; \mathcal{B})$, is defined by         
\begin{equation}
\hat{Q}\varphi(t) = \int_{0}^{T} A(t,\tau) \varphi(\tau ) \, d\tau = \int_{0}^{t} A(t,\tau)\varphi(\tau) \, d\tau ,
\end{equation}
where $\varphi \in \mathcal{B}$.  
Then, the Volterra integral equation with the  kernel $A(t,\tau)$ can be solved by successive approximations. That is, the
Neumann series converges in the topology of $L^{1 }(I;\mathcal{B} )$.
\end{theorem}
\goodbreak\pf
Let $  L^{ 1 }( J  ;\mathcal{B} )$ be the Banach space with norm $ \| \cdot \|_{L^{ 1  }( J ; \mathcal{B})}$, where $J = (0, t )$ and  where $0 < t \leq T <\infty$. Suppose the function $ \psi_{0} : J \to L^{ 1 }( J  ;  \mathcal{B})$ is a bounded function with norm 
\begin{equation}
\| \psi_{0} \|_{ L^{1}( J ; \mathcal{B})} =  \int_{0}^{t} \|\psi_{0}(\tau)\|  \, d\tau
\end{equation}
Define the Volterra integral operator in the following way,
\begin{equation}
 \hat{Q}\phi(t) = \int_{0}^{t} A(t,\tau)\phi(\tau) \, d\tau ,
\end{equation}
where $\phi \in \mathcal{B}$. 
Suppose that $\forall (t , \tau) \in \bar{I}^{2}$, $A(t, \tau): \mathcal{B} \to \mathcal{B}$ is a measurable and uniformly bounded operator
. Then by definition 6, $\exists D>0$ such that $\forall (t, \tau) \in  \bar{I}^{2}$, $\| A(t, \tau) \| = D < \infty$.
Furthermore, $A(t,\tau) = 0 $ when $\tau > t$. 
Then by lemma \ref{LemmaV3}, it follows that the norm of $\psi_{0} (t) $ satifies the condition $\| \psi_{0} \|_{L^{  1 }( J ;\mathcal{B} )}  \leq N  t^{0} = N $, since $n = 0$. Then, we compute $L^{1}$ norm estimates for the following equation:
\begin{eqnarray}
\fl{
\eqalign{
\| \psi_{1}  \|_{L^{ 1  }( J ;  \mathcal{ B } ) } &= \|\hat{Q} \psi_{0} \|_{  L^{  1  }(  J;  \mathcal{ B } ) }  = 
\int_{0}^{t}  \biggl \| \int_{0}^{t_{1}   }  A( t_{1}, \tau ) \psi_{0}  (\tau) \, d\tau   \biggr \|   \, d t_{1  }  \\
& \leq 
\int_{0}^{ t }  
\int_{0}^{ t_{1}  } \|A(t_{1} ,\tau)\| \, \|\psi_{0}(\tau) \|    \, d\tau    d t_{1  }     \leq 
 \int_{0}^{  t }  \int_{0}^{  t_{1}   }  D  \|\psi_{0} (\tau) \| \, d\tau   t_{1}  \\
 &
  \leq 
 \int_{0}^{  t }   D  \| \psi_{0} \|_{L^{  1  }(  J ;  \mathcal{B} ) }  \, d t_{1  }  
\leq 
D N \int_{0}^{t}  \, d   t_{1  }     = D N t 
}
}
\end{eqnarray}
where $ t_{1}  \leq t$ and if $ N = \| \psi_{0}  \|_{L^{  1  }(  J ;  \mathcal{B} ) }$, then if follows that 
\begin{equation} \label{bdestimate}
\|\hat{Q} \psi_{0} \| _{L^{1}( J ;  \mathcal{B} ) }    \leq   D  \| \psi_{0} \| _{L^{ 1 }(  J ;  \mathcal{B} ) }  t.
\end{equation}
Then by inductively applying lemma \ref{LemmaV1}, we see that the $n$th term of the Neumann series $\psi_{n}$ gives the simplex structure:
\begin{equation}
\| \psi_{n} \|_{L^{ 1 }( J ; \mathcal{B} )}  \leq  D^{n} \|  \psi_{0} \|_{L^{ 1 }( J; \mathcal{B} )}  \frac{t^{n}}{n !} 
\end{equation}
Thus the series $ \sum_{n=0}^{\infty}  \psi_{n}$ is majorized by
\begin{equation}
 \|  \psi_{0}  \|_{L^{ 1 }( J; \mathcal{B} )} +  \|  \psi_{0}  \|_{L^{ 1 }( J; \mathcal{B} )}  \sum_{n=1}^{\infty}  D^{n}   \frac{t^{n}}{n} =  \|  \psi_{0} \|_{L^{ 1 }(  J ; \mathcal{B} ) } e^{D t}
\end{equation}
and the above estimate holds for all $t \in (0, T]$. 
Therefore, the Neumann series converges
with respect to the topology  $L^{  1 } (I;\mathcal{B} )$.  \quad $ \blacksquare $\\

\begin{theorem} \label{LpVolterra}($L^{p}$ Volterra Theorem)
 Let $I = (0, T) $ such that $0 < T \leq \infty$. Assume  that the kernel $A(t,\tau)$ is a measurable and uniformly bounded linear operator (in the sense of definition 6).  Furthermore, assume that
$\forall (t, \tau) \in \bar{I}^{2} $,  $A(t, \tau): \mathcal{B} \to \mathcal{B}$ where $\mathcal{B}$ is a Banach space. Suppose that the kernel satisfies the following condition, $A(t,\tau ) = 0$, when $ \tau > t$.  The Volterra integral operator, $\hat{Q}: L^{ p  }( I ;\mathcal{B} ) \to L^{p }(I ; \mathcal{B})$, is defined by         
\begin{equation}
\hat{Q}\varphi(t) = \int_{0}^{T} A(t,\tau) \varphi(\tau ) \, d\tau = \int_{0}^{t} A(t,\tau)\varphi(\tau) \, d\tau ,
\end{equation}
where $\varphi \in \mathcal{B}$.  
Then, the Volterra integral equation with the 
above kernel $A(t,\tau)$ can be solved by successive approximations. That is, the
Neumann series converges in the topology of $L^{p }(I;\mathcal{B} )$.
\end{theorem}
\goodbreak\pf
Let $  L^{ p }( J  ;\mathcal{B} )$ be the Banach space with norm $ \| \cdot \|_{L^{ p }( J ; \mathcal{B})}$, where $J = (0, t )$ and  where $0 < t \leq T <\infty$. Suppose the function $ \psi_{0} : J \to L^{ p }( J  ;  \mathcal{B})$ is a bounded function with norm 
\begin{equation}
\| \psi_{0} \|_{ L^{ p }( J ; \mathcal{B})} =  \biggl ( \int_{0}^{t} \|\psi_{0}(\tau)\|^{p}  \, d\tau \biggr )^{1/p}
\end{equation}
Define the Volterra integral operator in the following way,
\begin{equation}
 \hat{Q}\phi(t) = \int_{0}^{t} A(t,\tau)\phi(\tau) \, d\tau ,
\end{equation}
where $\phi \in \mathcal{B}$. 
Suppose that $\forall (t , \tau) \in \bar{I}^{2}$, $A(t, \tau): \mathcal{B} \to \mathcal{B}$ is a measurable and uniformly bounded operator
. Then by definition 6, $\exists D>0$ such that $\forall (t, \tau) \in  \bar{I}^{2}$, $\| A(t, \tau) \| = D < \infty$.
Furthermore, $A(t,\tau) = 0 $ when $\tau > t$. 
Then by lemma \ref{LemmaV2}, it follows that the norm of $\psi_{0} (t) $ satifies the following condition, $\| \psi_{0} \|_{L^{ p}( J ;\mathcal{B} )}  \leq N  t^{ 0} = N  $, since $n = 0$. 
The Banach space norm of the function $\hat{Q} \psi_{0}(t)$ is
\begin{eqnarray}
\fl{
\eqalign{
\|  \hat{Q}  \psi_{0}( t_{1} ) \|  & \leq  \biggl  (  \int_{0}^{t_{1} }  \| A(  t_{1}, \tau)  \| ) \,   d\tau   \biggr )^{1/q}  \biggl ( \int_{0}^{t_{1}  }  \| A(t_{1}  , \tau) \| \, \| \psi_{0}( \tau ) \|^{p}  \, d\tau \biggl )^{1/p}  \\
& \leq D^{1/q} \biggl  (  \int_{0}^{t_{1} }    \,   d\tau   \biggr )^{1/q}  \biggl ( \int_{0}^{  t_{1}  }  D \|  \psi_{0}( \tau ) \|^{p}  \, d\tau \biggl )^{1/p}  \\
&
\leq D^{1/q}   D^{1/p}   t_{1}^{1/q}  \biggl ( \int_{0}^{  t_{1}  }  \| \psi_{0} ( \tau ) \|^{p}  \, d\tau \biggl )^{1/p}  
\leq D t_{1 }^{1/q} \biggl ( \int_{0}^{  t_{1}  }  \| \psi_{0} ( \tau ) \|^{p}  \, d\tau \biggl )^{1/p}  .
}
}
\end{eqnarray}
Raising both sides to the $p$th power and integrating, we see by Fubini's theorem that
\begin{eqnarray}
\fl{
\eqalign{
 \int_{0}^{  t_{1}  }   \| \hat{Q}  \psi_{0}( t_{1})  \|^{p}  \, d t_{1}  &  \leq  \int_{0}^{   t }   D^{p}  t_{1}^{p/q}  \int_{0}^{  t_{1}  }  \| \psi_{0} ( \tau ) \|^{p}  \, d\tau d t_{1}  
\leq D^{p}    \int_{0 }^{ t  }   t_{1}^{p/q}     \| \psi_{0}  \|_{  L^{p }  ( J ; \mathcal{ B}  ) }^{p}  \,  d t_{1}   \\
&
 \leq D^{p}   \int_{0}^{t}    N^{p}  t_{1}^{p/q}     \, d t_{1}  
  \leq D^{p} N^{p}   \frac{ t^{p/q} + 1 }{  \frac{p }{ q } + 1 }     \\
 &  \leq D^{p}  N^{p}    \frac{  t^{  p/q  + 1 }  }{  p }      
}
}
\end{eqnarray}
where $ t_{1}  \leq t$. Now, taking the $p$th root, we obtain
\begin{equation} \label{bdestimate}
\|\hat{Q} \psi_{0} \| _{L^{p}( J ;  \mathcal{B} ) }    \leq   D N \frac{  t^{1/q  + 1/p}}{p ^{1/p}} = \frac{ D N }{ p^{1/p}  }  t 
\end{equation}
and if $ N = \| \psi_{0}  \|_{  L^{ p } (  J ;  \mathcal{B} ) }$, then if follows that 
\begin{equation} \label{bdestimate}
\|\hat{Q} \psi_{0} \| _{L^{p}( J ;  \mathcal{B} ) }    \leq   \frac{ D }{p^{1/p} } \| \psi_{0}  \|_{  L^{ p } (  J ;  \mathcal{B} ) }   t 
\end{equation}
converges with respect to the norm $\| \cdot \|_{L^{p }  }$.
Then by inductively applying lemma \ref{LemmaV2}, we see that the $n$th term of the Neumann series $\psi_{n}$ gives the following result:
\begin{equation}
\| \psi_{n} \|_{L^{ p }( J ; \mathcal{B} )}  \leq \frac{ D^{n} }{ p ^{n/p}} \|  \psi_{0} \|_{L^{ p }( J; \mathcal{B} )}   \frac{t^{n}   }{  (n ! )^{1/p} }   
\end{equation}
Thus the series $ \sum_{n=0}^{\infty}  \psi_{n}$ is majorized by
\begin{equation}
 \|  \psi_{0}  \|_{L^{ p }( J; \mathcal{B} )} +  \|  \psi_{0}  \|_{L^{ p }( J; \mathcal{B} )}  \sum_{n=1}^{\infty}  \frac{ D^{n}  }{ p ^{n/p }  }   \frac{  t^{n}   }{ (  n  !  )^{1/p} }   
\end{equation}
Now, we investigate whether the Neumann series are convergent. The ratio test will be used to determine whether the infinite series converges or diverges. Let $L$ be defined in the following manner, 
\begin{equation}\label{L}
L = \lim_{n \to \infty}   \biggl | \frac{ a_{n+ 1 }  }{ a_{n} } \biggr |
\end{equation}
In this case, $a_{n} = \|  \psi_{n}  \|_{L^{ p }( J; \mathcal{B} )}  $, and $a_{n+ 1} =  \|  \psi_{n+ 1}  \|_{L^{ p }( J; \mathcal{B} )} $. Since  $\|  \psi_{n}  \|_{L^{ p }( J; \mathcal{B} )} $ is always a positive number, the absolute values in equation \eref{L} can be removed and we obtain: 
\begin{eqnarray}
\fl{
\eqalign{
L  & = \lim_{n \to \infty }   \frac{   \|  \psi_{n+ 1}  \|_{L^{ p }( J; \mathcal{B} )}  }{   \|  \psi_{n}  \|_{L^{ p }( J; \mathcal{B} )} }  
= 
\lim_{n \to \infty }  
  \frac{ \|  \psi_{0}  \|_{L^{ p }( J; \mathcal{B} )}    M(p)^{n + 1  }  t^{n + 1 }          }{  [ ( n + 1) !]^{1/p} }  \cdot 
    \frac{   ( n ! )^{1/p}   }{  \|  \psi_{0}  \|_{L^{ p }( J; \mathcal{B} )}  M(p)^{n }  t^{n }   }  \\
    & =   M(p)  t   \lim_{n \to \infty }     
         \frac{  1  }{  ( n  +1 )^{1/p} } = 0 
}
}
\end{eqnarray}
where, $M(p ) = \frac{  D}{ p^{1/p} }$.
Thus $L = 0 <1 $, and by the Ratio Test theorem, the Neumann series converges absolutely. Therefore, the
Neumann series converges in the topology of $L^{p }(I;\mathcal{B} )$. \qquad $ \square $\\

\section{Applications of the Volterra Theorem} \label{section4}
In this section, we will present several different types of applications of theorems
\ref{volterratheorem} and \ref{L1Volterra}. The first example is classical and the second example is the unitary quantum-mechanical example. The closest example to quantum mechanics is example 2 where the spatial operator is
a unitary operator. Each example presents two versions, corresponding to the $L^{\infty}$ Volterra and $L^{p}$ Volterra theorems, respectively.
The following examples are not new and are well-known in the literature. These examples serve as quick applications of the Volterra theory presented in section \ref{section4}, with serious applications delayed to later papers

Let $I$ be an interval in the temporal dimension such that $I = (0, T)$ for some $T>0$. In the following examples, the time interval $I$ will always be the same time interval $(0, T)$. Furthermore, let $L^{n,m}(I; \mathbb R^{d})$ be the
Banach space of $L^{m}(\mathbb R^{d})$ functions over $I$. Thus we will define the Lebesgue space
$L^{n,m}(I; \mathbb R^{d})$ as
\begin{equation}\fl
L^{n,m}(I; \mathbb R^{d}) = \biggl \{  \phi :
 \biggl ( \int_{I} \biggl [ \int_{\mathbb R^{d}} |\phi(y,\tau)|^{m} \, dy \biggr ]^{n/m} \, d\tau \biggr  )^{1/n} = \| \phi \|_{L^{n,m}(I ; \mathbb R^{d})} < \infty \biggr \}.
\end{equation}
If $m$ and $n$ are equal, then the Lebesgue space $L^{n,m}(I ; \mathbb R^{d})$ will be written as $L^{n}(I; \mathbb R^{d})$. 

\subsection{Example 1} 
\subsubsection*{$L^{\infty}$ Case:}
Let the Banach space $\mathcal{B}$ be $ L^{2}(\mathbb R^{n})$ and consider a bounded integrable (e.g., continuous) real or complex-valued kernel $A(t,\tau)$, satisfying the Volterra condition in $(t,\tau)$.   The Hilbert-Schmidt kernel is a
function $K: \mathbb R^{n} \times \mathbb R^{n} \to \mathbb{F}$ on the space variables, where $ \mathbb{F} = \{ \mathbb{C}, \mathbb{R} \} $. The norm of the Hilbert-Schmidt kernel is given by
\begin{equation}
\biggl ( \int_{\mathbb R^{n} \times \mathbb R^{n}} |K(x,t;y,\tau)|^{2} \, dx dy \biggr )^{1/2} = \| K(t, \tau) \|_{L^{2}(\mathbb R^{2n})} \leq N < \infty 
\end{equation}
The linear operator $A(t,\tau)$ is defined on $L^{\infty}(I^{2})$, and $A(t,\tau)$ is a Hilbert-Schmidt operator. Then the Hilbert-
Schmidt operator $A(t,\tau) :L^{2}(\mathbb R^{n}) \to L^{2}(\mathbb R^{n})$ is given by
\begin{equation}
 A(t,\tau)\phi(t)  = \int_{\mathbb R^{n}} K(x,t;y,\tau) \phi(y, \tau)  \, dy \quad  \forall \phi \in L^{\infty, 2}(I ; \mathbb R^{n}) 
\end{equation}  
It follows that the operator $A(t,\tau)$ is bounded.
The function $K(x,t;y,\tau)$ belongs to $L^{\infty,2}(I^{2} ; 
\mathbb R^{2n})$. \
Then we take the absolute values of $A(t,\tau) \phi(t)$ and we obtain
\begin{eqnarray}
|A(t,\tau) \phi(t)| & \leq \int_{\mathbb R^{n}} |K(x,t;y,\tau)| 
|\phi(y,\tau)| dy \nonumber \\
& \leq \biggl ( \int_{\mathbb R^{n}} |K(x,t;y,\tau)|^{2} dy \biggr)^{1/2}  \biggl  ( \int_{\mathbb R^{n}} |\phi(y,\tau)|^{2} dy \biggr) ^{1/2}
\end{eqnarray}
and hence,
\begin{equation}
\|A(t,\tau) \phi(t)\|_{L^{2}(\mathbb R^{n})} \leq  
 \| K(t, \tau) \|_{L^{2}(\mathbb R^{2n})}  
\|\phi(\tau) \|_{L^{2}(\mathbb R^{n})} \leq N \|\phi(\tau) \|_{L^{2}(\mathbb R^{n})} 
\end{equation}
with
\begin{equation}
N \equiv \| K \|_{L^{\infty, 2}(I^{2} ; \mathbb R^{2n})}
\end{equation}
Therefore by the $L^{\infty}$ Volterra Theorem, the Volterra integral equation with
a Hilbert-Schmidt kernel $K(x,t;y,\tau) \in L^{\infty,2}(I^{2}; \mathbb R^{n})$ 
can be solved by successive approximations.

\subsubsection{$L^{1}$ Case:}

Now we will provide an example for the $L^{1}$ Volterra Theorem. 
The difference between this example and the previous one is that the 
Lebesgue space in time is $L^{1}(I)$ rather than $L^{\infty}(I)$.
We will assume the same hypotheses for the Volterra kernel $A(t, \tau) $ and the kernel $K(x, t; y, \tau)$  as in the $L^{\infty}$ case.
Thus,
\begin{equation}
\| A(t,\tau) \phi(t) \|_{L^{2}(\mathbb R^{n})} \leq  D \| \phi(\tau) \|_{L^{2}(\mathbb R^{n})}
\end{equation}
and hence,
\begin{equation}
\| A \phi \|_{L^{1}(I ; \mathbb R^{n})} \leq  \int_{0}^{t}    D   \| \phi  (\tau ) \|_{L^{2}( \mathbb R^{n})} \, d\tau = D  \| \phi   \|_{L^{1,2}( I ;  \mathbb R^{n})} 
\end{equation}
Thus, we have shown that the norm of $A(t,\tau) \phi(t)$ is bounded, and hence
\begin{equation}
\| \psi \|_{L^{1}(I ; \mathbb R^{n})}  \leq   \int_{0}^{t}     D \| \phi  \|_{L^{1,2}(I ; \mathbb R^{n})}  \, d\tau =   D \| \phi  \|_{L^{1,2}(I ; \mathbb R^{n})}  t 
\end{equation}
Therefore by the $L^{1}$ Volterra Theorem, the Volterra integral equation with
a Hilbert-Schmidt kernel in space and a uniformly bounded kernel in time can be solved by successive approximations.

\subsection{Example 2}
\subsubsection{$L^{\infty}$ Case:} Let $V(x,t)$ be a bounded potential, and $x \in \mathbb R^n$.
The potential $V$ may be time-dependent, but in that case its bound should be 
independent of $t$ (i.e., $V \in L^{\infty}(I ; \mathbb R^{n})$, with $
\|V\|_{L^{\infty}(I ; \mathbb R^n)} \equiv C$).
Let the Banach space $\mathcal{B}$ be the Hilbert space $L^{2}(\mathbb R^n)$.
Recall that  the solution $u(t)$ is given by the Poisson integral formula
\begin{equation}
u(t) \equiv \int_{\mathbb R^{n} } K_{f}(x, y, t) h(y) \, dy 
\end{equation}
where $K_{f}(x,y,t) = (4 \pi i t)^{-n/2} e^{i |x-y|^{2}/4 t}$,
is the solution of the free Schr\"{o}dinger equation with initial data 
$u(x,0) = h(x)$ in $L^{2}(\mathbb R^n)$. 

It is well known that $U_{f}(t,\tau)$ is  unitary, 
and hence the norm of $U_{f}$ as an operator from $L^{2}(\mathbb R^n)$ to itself is 
$\|U_{f}(t,\tau)\|_{L^{2}(\mathbb R^{2n})} =1$. A proof that the operator $U_{f}(t,\tau)$ is a unitary operator can be found in \cite[~Ch.4]{Evans}.
We wish to solve the Schr\"{o}dinger equation with the potential $V$ by iteration.
The Volterra integral equation is given by
\begin{equation} \label{integralvolterra}
u(x,t) + i\int_{0}^{t} \hat{U}(t-\tau) V(\tau) u(\tau)\, d\tau  = \hat{U} f(x) 
\end{equation} 
where, 
\begin{equation} \label{UVequation}
\hat{U}(t-\tau)V(\tau) u(\tau) = \int_{\mathbb R^{n}} K_{f}(x,t;y,\tau) V(y,\tau) u(y,\tau) \, dy 
\end{equation}
Hence, the Volterra theorem applies.  

In theorem \ref{volterratheorem}, take $\mathcal{B} = L^{2}(\mathbb R^n)$, $A = UV$ as defined in equation \eref{UVequation}. It remains to check that $UV$ 
is a bounded operator on $L^{2}(\mathbb R^n)$ with bound independent of $t$ and $\tau$.
Here $V(\tau)$ is the operator from $L^{2}(\mathbb R^n)$ to $L^{2}(\mathbb R^n)$  defined by multiplication 
of $f(y,\tau)$ by $V(y,\tau)$, and $\|V(\tau)\|$ is its operator norm.  But
\begin{eqnarray}
\|V(\tau)f(\tau) \|_{L^{2}(\mathbb R^{n})}^2 &= \int_{\mathbb 
R^{n}} |V(y,\tau) f(y,\tau)|^2 \,  dy \nonumber \\
&\le
C^2 \int_{\mathbb R^{n}} |f(y,\tau)|^{2} \, dy 
= C^2 \|f(\tau) \|_{L^{2}(\mathbb R^{n})}^2.
\end{eqnarray}
Therefore,
\begin{equation}
\|V(\tau) f(\tau)\|_{L^{2}(\mathbb R^{n})} \le  C \|f(\tau)\|_{L^{2}(\mathbb R^{n})} \quad \forall f \in L^{2}(\mathbb R^n).
\end{equation}
In other words $\|V \|_{L^{\infty}(I ; \mathbb R^{n})}$, the norm of the operator $V(\tau) \le
C \equiv \|V\|_{L^{\infty}(I ; \mathbb R^n) }$, is the uniform norm of the function 
$V(x,t)$.
Therefore,
\begin{equation}
\|U(t,\tau)V(\tau)f(\tau) \|_{L^{2}(\mathbb R^{n})} \le C \|f(\tau)\|_{L^{2}(\mathbb R^{n})}.
\end{equation}
and the operator norm  of $A = UV$ is bounded by $\|U(t,\tau)V(\tau) \|_{L^{2}(\mathbb R^{n})} \le C$.
Then,
\begin{equation}\fl
A(t,\tau) f(\tau) = \int_{\mathbb R^{n}} K(x,t;y,\tau) f(y, \tau) \, dy = \int_{\mathbb R^{n}} K_{f}(x,t;y,\tau) V(y,\tau)   f(y, \tau) \, dy
\end{equation}
Therefore, we obtain the following $ L^{\infty,2}$ norm estimate for $\hat{Q}f=  SVf $:
\begin{equation}
\| \psi \|_{L^{\infty, 2}(I ; \mathbb R^{n})} = \| SV f\|_{L^{\infty,2}(I ; \mathbb R^n)}  \leq  C \|f \|_{L^{\infty, 2}( I ; \mathbb R^{n})} T
\end{equation}
where
\begin{equation}
\psi(t) = SVf(t) = \int_{0}^{t} U(t,\tau)V(\tau) f(\tau) \, d\tau 
\end{equation}
Thus we have verified all the hypotheses of the Volterra Theorem, and we 
conclude that the solution of the Schr\"{o}dinger equation with potential 
$V$ is the series $\varphi = \sum_{n=0}^\infty \psi_n$, 
where $\psi_0(t) = f(t) = \hat{U}(t,\tau) h(x)$, and where $h(x)$ is the initial data.

\subsubsection{$L^{1}$ Case:}
Now we will consider an application for the $L^{1}$ Volterra Theorem. The difference between this example and the previous case is that the Lebesgue space in time is $L^{1} (I)$ rather than $L^{\infty}(I)$.
We will assume the same hypotheses for the functions $V$  and $f$ as in the $L^{\infty}$ case. Thus the norm of the potential function $V$ and $f$ is shown to be bounded and the inequality is given by
\begin{equation}
\|V(\tau) f(\tau)\|_{L^{2}(\mathbb R^{n})} \leq  C \|  f(\tau)\|_{L^{2}(\mathbb R^{n})} \quad \forall f \in \mathcal{H}.
\end{equation}
Then, we take the $L^{1}$ norm with respect to the time variable and we obtain 
\begin{equation}
\|V f \|_{L^{1, 2 }(I ; \mathbb R^{n} )} \leq C  \int_{0}^{t}   \|f \|_{L^{2}( \mathbb R^{n})} \, d\tau = C \| f \|_{L^{1, 2}( I; \mathbb R^{n})} 
\end{equation}
Also, the operator $A(t, \tau)=U(t, \tau) V(\tau )$ is bounded by $\| U(t, \tau)  V(\tau) \|_{L^{2}(\mathbb R^{n})} \leq C$.
Thus, we have shown that the norm of $V f$ is bounded, and hence
\begin{eqnarray}
\fl{
\eqalign{
\| \psi \|_{L^{1, 2}(I ; \mathbb R^{n})}  & =  
\int_{0}^{t}   \| \psi(\tau) \|_{L^{2}( \mathbb R^{n})}  \, d\tau = 
\int_{0}^{t}  \biggl \| \int_{0}^{t_{1}}  U(t_{1}, \tau) V(\tau) f(\tau) \, d\tau \biggl \|_{L^{2}( \mathbb R^{n})}  \ \, d t_{1}  
\\
& \leq  \int_{0}^{t} \int_{0}^{t} \| U(t, \tau) V(\tau)  \|_{ L^{2}( \mathbb R^{n})}  \ \| f(\tau) \|_{ L^{2}( \mathbb R^{n} )  }    \, d\tau  dt_{1} \\
& \leq    \int_{0}^{t} \int_{0}^{t_{1}  }  C  \| f(\tau) \|_{ L^{2}( \mathbb R^{n} )  }   \, d\tau  dt_{1} 
=  \int_{0}^{ t}  C    \| f  \|_{L^{1, 2}( I; \mathbb R^{n})}  \, dt_{1}
\\
& = C  \| f  \|_{L^{1, 2}(  I ; \mathbb R^{n} ) }  t 
}
}
\end{eqnarray}
where $t_{1} \leq t$. 
Therefore by the $L^{1}$ Volterra Theorem, the Volterra integral equation with
a unitary operator in space can be solved by successive approximations in the topology $L^{1, 2} ( I; \mathbb R^{n}  )   $. 

\section{Conclusion}
The similarities between
the  Schr\"{o}dinger equation and the heat equation were used to create a theoretical
framework which will give the solution to the Schr\"{o}dinger problem. The   $L^{\infty}$ Volterra theorem proves that Volterra integral equation with a uniformly bounded kernel can be solved by successive approximations with respect to the topology $ L^{\infty}(I; \mathcal{B}) $. The $L^{p}$ Volterra theorem  proves the more general case when $ L^{p}(I; \mathcal{B}) $, and where $1 < p < \infty$, but it does not prove the special case when $ p =1$. As it turns out, the $p = 1$ case is also proven by inductively applying  the $L^{1}$ lemma. Thus, the $L^{1}$ Volterra theorem proves that the Volterra integral equation with a uniformly bounded kernel can also be solved by successive approximations with respect to the topology $ L^{1}(I; \mathcal{B}) $

In future work I shall apply the  Volterra theorem in contexts more complicated than the simple examples 
presented here.  Preliminary work on these applications appears in 
Chapters 8 and 9 of the thesis \cite{merathesis}.  First, I hope to implement an idea 
due to Balian and Bloch \cite{BB74} to use a semiclassical propagator to construct 
a perturbation expansion for a smooth potential $V(x,t)$.   The solution of the Schr\"{o}dinger equation is given in terms of classical paths, and the semiclassical propagator $G_{scl} =  Ae^{iS/ \hbar}$ to the Green function is considered as the building block for the exact Green function \cite{BB74}. To prove convergence of the resulting 
semiclassical Neumann series under suitable technical conditions, in \cite[~Ch.8]{merathesis} a Semiclassical Volterra Theorem has been proved.  There is still more work to be done with regards to applying the Semiclassical Volterra theorem to various types of potentials. An application of the Semiclassical Volterra theorem is the potential problem in $\mathbb R^{n}$ considered in \cite{fuldar}.

The Volterra method will also be applied to the boundary value problem for the Schr\"odinger equation. The double-layer Schr\"{o}dinger operator will be shown to be bounded from a suitable space of functions defined on the boundary $ I \times \partial U$ to itself. At this stage, the boundary value problem remains unresolved.

\ack
I would like to acknowledge Dr.\ Stephen Fulling for guiding me in the right direction. He was always able to help me find an article or book which was essential to my thesis, and hence to the creation of this article. I would like to thank Dr.\ Fulling for his comments on the Volterra and General Volterra Theorems. His experience on functional analysis made it possible to create rigorous proofs. Another person who helped to generalize the assumptions of the Volterra theorems was Dr.\  Peter Kuchment. The most general technical assumptions of the Volterra Theorems were detected by Dr. Kuchment. 

Also,  I would like to give credit to Dr.\ Ricardo Estrada for his remarks and comments on the unitary operators.
Finally, I would alo like to thank Dr.\ Tetsuo Tsuchida, a visiting faculty member, for being willing to listen to several of my lectures regarding my thesis. I would also like to mention that Dr.\ Tsuchida was the professor who told me what book to read on the topic of divergent integrals and Abel summability. I would also like to say, that without this reference, my progress would have been substantially slowed down. 

This article was supported by the National Science Foundation Grants Nos. PHY-0554849 and PHY-0968269.

 \end{document}